\documentclass[11pt]{article}
\pdfoutput=1
\usepackage[nosort]{cite}
\usepackage{hyperref}
\hypersetup{colorlinks=true}
\hypersetup{linkcolor=black}
\hypersetup{citecolor=black}
\hypersetup{urlcolor=black}
\usepackage{geometry}
\usepackage{mathptmx}

\usepackage{epsfig}
\usepackage{amsmath,amssymb}
\usepackage{cite}

\renewcommand{\title}[1]{\vbox{\center\LARGE{#1}}\vspace{5mm}}
\renewcommand{\author}[1]{\vbox{\center#1}\vspace{5mm}}

\usepackage{slashed}

\usepackage{clay}

\usepackage{graphicx,color,subfig}

\usepackage{mciteplus}

\date{May 2022}
\title{Snowmass White Paper: Generalized Symmetries in Quantum Field Theory and Beyond}

\institution{Chicago}{\centerline{${}^{1}$Enrico Fermi Institute and Kadanoff Center for Theoretical Physics, University of Chicago}}
\institution{UCLA}{\centerline{${}^{2}$Mani L. Bhaumik Institute for Theoretical Physics, UCLA}}
\institution{UCSD}{\centerline{${}^{3}$Department of Physics, University of California, San Diego}}
\institution{SBU}{\centerline{${}^{4}$C.N. Yang Institute for Theoretical Physics, Stony Brook University}}

\authors{Clay C\'ordova\worksat{\Chicago}\footnote{e-mail: {\tt clayc@uchicago.edu}},
Thomas T. Dumitrescu\worksat{\UCLA}\footnote{e-mail: {\tt tdumitrecu@physics.ucla.edu}},
Kenneth Intriligator\worksat{\UCSD}\footnote{e-mail: {\tt keni@ucsd.edu}},
Shu-Heng Shao\worksat{\SBU}\footnote{e-mail: {\tt shu-heng.shao@stonybrook.edu}}
}

\begin{document}
\abstract{\noindent Symmetry plays a central role in quantum field theory.  Recent developments include symmetries that act on defects and other subsystems, and symmetries that are categorical rather than group-like. These generalized notions of symmetry allow for new kinds of anomalies that constrain dynamics.  We review some transformative instances of these novel aspects of symmetry in quantum field theory, and give a broad-brush overview of recent applications.}

\maketitle

\tableofcontents

\section{Introduction}

Symmetry has long been a transformative, development-guiding  notion in physics.  Examples include Noether's theorems \cite{Noether:1918zz,Byers:1998bd} connecting symmetries and conservation laws, and Landau's symmetry-based theory of phase transitions \cite{Landau:1937obd}.\footnote{~This paper is one of Landau's famous contributions during his tenure at the Ukrainian Physico-Technical Institute in Kharkiv.}  Transitions between quantum states are constrained by the Wigner-Eckart theorem, which applies to quantum systems whose Hamiltonian is invariant under a symmetry group. Here we will discuss symmetry in the context of quantum field theory (QFT), where additional structures such as locality or Lorentz invariance lead to a rich set of refinements and generalizations of the notion of symmetry. The resulting generalized symmetries can act on lower-dimensional defects and subsystems within the QFT, and their properties are in general captured by higher categories rather than by groups. 

Given that symmetry is such a vast subject, this overview will necessarily be an incomplete sketch of a few recent, selected lines of development, with many omissions.  As advertised we primarily focus on symmetries in QFT, with emphasis on renormalization group (RG) flows and anomalies. The later are intimately related to the possibility of projectively realizing symmetries in quantum theory. For a different recent survey from a condensed matter perspective, see for instance~\cite{McGreevy:2022oyu}. 

The fundamental forces are described via local gauge interactions. In that context, gauge invariance reflects a convenient but unphysical redundancy introduced to satisfy the requirements of locality and Lorentz invariance in relativistic QFT; as such it is exact, but has no observable consequences, e.g.~it can famously not be spontaneously broken~\cite{Elitzur:1975im}. 

By contrast, global symmetries imply relations between observable physical quantities; as such they may be spontaneously broken, and they are useful even if only approximate. Examples include the approximate unbroken $SU(3)_\text{flavor}$ and $SU(2)_\text{isospin} $ symmetries of QCD, as well as the approximate discrete symmetries~$C$ and $P$.  QCD spontaneously breaks its  approximate  chiral quark flavor symmetries; the pions are the associated (pseudo-) Nambu-Goldstone bosons (NGBs). Such considerations led to the discovery of the Adler-Bell-Jackiw (ABJ) anomaly~\cite{Adler:1969gk,Bell:1969ts} \footnote{~It was recently pointed out that the ABJ anomaly actually leads to non-invertible global symmetries~\cite{Choi:2022jqy,Cordova:2022ieu}.} and the closely related explicit violation of ABJ-anomalous symmetries by instantons~\cite{tHooft:1986ooh}.\footnote{~See~\cite{Coleman:1985rnk} for an influential account of symmetry in the development of QFT.}  Global symmetries can powerfully constrain QFT dynamics, even in strongly coupled theories.  't Hooft noted in~\cite{ tHooft:1979rat} that anomalies involving only global currents (unlike the ABJ case) are obstructions to gauging and they must be constant on renormalization group (RG) flows; such anomalies are often now often referred to as {\it 't Hooft anomalies} and will be reviewed here. 

Global symmetries are termed accidental (or emergent) when they arise or enhance in the deep IR and violated by higher dimension operators. Examples are $U(1)_B$ baryon and $U(1)_L$ lepton number, which are preserved by the dimension 4 operators of the Standard Model (SM) but are violated by higher dimension operators (as well as by electroweak instantons).\footnote{~The combination~$U(1)_{B-L}$ is exactly preserved in the SM, but is expected to be violated by physics beyond the SM (BSM).}  All global symmetries are expected to be accidental in quantum gravity, see for instance the recent review\cite{Harlow:2022gzl} and references therein.  

There have been many attempts to generalize global symmetries in relativistic QFT, e.g.~by combining internal and Lorentz symmetries, or by contemplating conserved currents of higher spin~$s \geq 3$.\footnote{~These are symmetric tensors of the Lorentz group with~$s\geq 3$ indices. By contrast, the stress tensor~$T_{\mu\nu}$ is a two-index Lorentz tensor of spin~$s = 2$.} The latter famously arise in the context of integrable QFTs in~$d = 2$ spacetime dimensions, as in~\cite{Zamolodchikov:1978xm}.
For $d>2$, the Coleman-Mandula (CM) theorem~\cite{Coleman:1967ad} implies that higher spin conserved currents essentially render the theory free. 

There are two interesting, broad classes of exceptions to the CM theorem:  conformal field theories, where the Poincar\'e group is extended to the conformal group $SO(d,2)$,\footnote{~A CFT analogue of the CM theorem was proved in~\cite{Maldacena:2011jn}.} and supersymmetric field theories~\cite{Haag:1974qh}; these two classes intersect in the superconformal theories~\cite{Nahm:1977tg}. We will not specifically focus on these classes of theories here; they are covered in other Snowmass white papers (see e.g.~\cite{Argyres:2022mnu,Poland:2022qrs}). Instead, we will survey other large classes of generalized symmetries that are not ruled out by the CM-theorem and that require neither conformal symmetry nor supersymmetry (though they do occur in conformal and supersymmetric theories).  

The remainder of this white paper is organized as follows: 

In section \ref{sec:gensym} we review basic aspects of ordinary symmetries (section~\ref{sec:ord}) before surveying four kinds of generalized global symmetries: higher form symmetries (section~\ref{sec:hf}), higher group symmetries (section~\ref{sec:hg}), non-invertible symmetries (section~\ref{sec:noninv}) and subsystem symmetries (section~\ref{sec:fractons}). The first three generalizations can arise in relativistic systems, while the fourth one is only known to occur in non-relativistic contexts, notably in condensed matter models of fractons. 

Section \ref{sec:anomdyn} is concerned with 't Hooft anomaly matching and applications to QFT dynamics. In section~\ref{sec:anomalies} we review the notion of an 't Hooft anomaly for ordinary global symmetries, as well as the closely associated concepts of anomaly matching and anomaly inflow. We then outline how these anomalies extend to the generalized symmetries introduced in section \ref{sec:gensym}. In section \ref{sec:dynamics} we survey some recent applications of anomaly matching for ordinary and global symmetries.

\section{From Ordinary to Generalized Symmetries}\label{sec:gensym}

\subsection{Ordinary Symmetries} \label{sec:ord}

Here we summarize a few standard facts about ordinary global symmetries that will set the stage for subsequent generalizations.  Throughout this paper, $d$ stands for the spacetime dimension.

In a QFT with an ordinary, or 0-form, global symmetry~$G^{(0)}$, there is a topological defect~$U^{(0)}(g, \Sigma_{d-1})$ for every group element~$g \in G^{(0)}$ that is supported on a closed codimension-1 submanifold~$\Sigma_{d-1}$ of spacetime.\footnote{~We sometimes call~$U^{(0)}(g, \Sigma_{d-1})$ a symmetry defect. For an exposition of this perspective, see for instance~\cite{Gaiotto:2014kfa} and references therein. Throughout we use~$X^{(p)}$ to indicate that~$X$ is a $p$-form.} The fusion of the~$U^{(0)}(g, \Sigma_{d-1})$ is governed by the group law for~$G^{(0)}$.\footnote{~If~$G^{(0)}$ has an 't Hooft anomaly, it shows up in more complicated configurations of the symmetry defects (see for instance~\cite{Gaiotto:2014kfa}).} An elementary example is~$G^{(0)} = U(1)$, in which case Noether's theorem implies a conserved 1-form current~$j^{(1)}$,
\begin{equation}
d * j^{(1)} = 0~.
\end{equation}
We can then construct a conserved charge~$Q^{(0)}(\Sigma_{d-1})$, whose exponential gives the~$U(g, \Sigma_{d-1})$,
\begin{equation}
Q^{(0)}(\Sigma_{d-1}) = \int_{\Sigma_{d-1}} * j^{(1)}~, \qquad U^{(0)}(g = e^{i \alpha}, \Sigma_{d-1}) = \exp\left(i \alpha Q^{(0)}(\Sigma_{d-1})\right)~.
\end{equation}
Current conservation and Stokes' theorem imply that the dependence on~$\Sigma_{d-1}$ is topological, as long as the deformation of~$\Sigma_{d-1}$ does not collide with other insertions. 
 The group multiplication law follows from the additivity of the charge. In the most familiar setup, $\Sigma_{d-1}$ is a spatial slice in Lorentzian signature; in Euclidean signature~$\Sigma_{d-1}$ can be any suitable closed codimension-1 cycle. Some comments are in order:
\begin{itemize}
\item If~$G^{(0)}$ is non-Abelian and connected, the group law for the~$U^{(0)}(g, \Sigma_{d-1})$ can be deduced from the fact that the OPE of the currents~$j_a^{(1)}$ encodes the structure constants~$f_{abc}$ of the Lie algebra of~$G^{(0)}$, 
\begin{equation}
j_a^{(1)} (x) j_b^{(1)}(0) \sim C(x) f_{abc} j_c^{(1)}(0) + \cdots~,
\end{equation}
with~$C(x)$ a known, singular function depending on the Lorentz indices of the currents (which we suppress). 
\item If~$G^{(0)}$ is not connected (e.g.~it could be a discrete group) then some of the symmetry defects~$U^{(0)}(g, \Sigma_{d-1})$ are not associated with conserved currents; nevertheless they typically exist and have the properties outlined above. This is a generalization of Noether's theorem.\footnote{~See for instance~\cite{Harlow:2018tng} for a recent discussion of possible failure modes, with references.} 

\item $G^{(0)}$ naturally acts on local operators by linking them with~$U^{(0)}(g, \Sigma_{d-1})$. These operators in turn create states (e.g.~point particles) charged under~$G^{(0)}$. More generally, $G^{(0)}$ acts on the Hilbert space~$\cal H$ in Lorentzian signature. 

\item If the vacuum is preserved by a subgroup~$H^{(0)} \subset G^{(0)}$, then the $G^{(0)}$ is spontaneously broken to~$H^{(0)}$. This leads to a vacuum manifold~${\cal M} = G^{(0)} / H^{(0)}$, and hence to Nambu-Goldstone (NG) bosons (if~${\cal M}$ has positive dimension) or domain walls between the disconnected components of~${\cal M}$. The Hilbert space~$\cal H$ in each vacuum furnishes a representation of~$H^{(0)}$. 

\item A QFT$_d$ with~$G^{(0)}$ symmetry can be coupled to a non-dynamical background~$G^{(0)}$ connection~$A^{(1)}$ in a way that preserves background gauge invariance. If~$G^{(0)}$ has an 't Hooft anomaly this requires placing the theory at the boundary of an invertible anomaly theory~${\cal A}_{d+1}[A^{(1)}]$ that depends on the~$G^{(0)}$ connection.\footnote{~If one insists on a strictly~$d$-dimensional description, the anomaly spoils background gauge-invariance in a mild and familiar way, e.g.~the partition function is multiplied by complex phases under~$G^{(0)}$ gauge transformations.} 
The symmetry can then be gauged by path-integrating over~$A^{(1)}$. 
\end{itemize}

\subsection{Higher Form Symmetries}\label{sec:hf}

CM-like theorems famously limit possibly generalizations of~$G^{(0)}$ symmetry in relativistic QFT (see introduction; see below for a discussion of  non-relativistic systems). One such generalization, suggested by experience from string theory and SUSY QFT,\footnote{~In string theory, $p$-form symmetries (like all symmetries) are naturally gauged. In SUSY QFT's they can be global symmetries, and the associated charges can extend the SUSY algebra (see~\cite{Dumitrescu:2011iu} and references therein).} was spelled out in \cite{Kapustin:2014gua,Gaiotto:2014kfa}:\footnote{~See~\cite{Sharpe:2015mja,Gaiotto:2017yup,Hofman:2017vwr,Hsin:2018vcg,Hsin:2019gvb,Morrison:2020ool,DelZotto:2020esg,Gukov:2020btk} for a highly incomplete sampling of later work focusing on higher-form symmetries.} replace the 1-form current~$j^{(1)}$ by a~$(p+1)$-form current~$J^{(p)}$ satisfying
\begin{equation}
d * J^{(p+1)} = 0~.
\end{equation}
As before we can construct a charge~$Q^{(p)} = \int_{\Sigma_{d - p - 1}} * J^{(p+1)}$ by integrating over a codimension-$(p+1)$ cycle, and a symmetry defect~$U^{(p)}(g = e^{i\alpha}, \Sigma_{d - p - 1}) = \exp(i \alpha Q^{(p)}(\Sigma_{d - p - 1}))$, both of which depend topologically on~$\Sigma_{d- p -1}$. The symmetry defects satisfy a~$U(1)$ group law, and we say that they generate a~$U(1)$ $p$-form symmetry, denoted~$U(1)^{(p)}$. An important example is free Maxwell theory in any spacetime dimension~$d$, based on a dynamical~$U(1)$ connection~$a^{(1)}$ with field strength~$f^{(2)} = da^{(1)}$. The Maxwell equations,
\begin{equation}
d * f^{(2)} =0,~~ d f^{(2)}= 0~,
\end{equation}
can be read as conservation equations for an electric~$2$-form current~$f^{(2)}$ and a magnetic ~$(d - 2)$-form current~$* f^{(2)}$. Thus the theory has~$U(1)^{(1)}_e  \times U(1)^{(d-3)}_m$ symmetry.\footnote{~The statement that these are compact~$U(1)$ symmetries is related to the fact that~$a^{(1)}$ is a~$U(1)$ connection, with quantized electric charges and magnetic fluxes. The~$U(1)^{(1)}_e$ and~$U(1)^{(d-3)}_m$ symmetries have a mixed 't Hooft anomaly~\cite{Gaiotto:2014kfa} that plays an important role in many applications.} 
Some comments are in order:

\begin{itemize}
\item In analogy with 0-form symmetries, we can also consider discrete~$(p \geq 1)$-form symmetries~$G^{(p)}$, which are necessarily abelian.\footnote{~This is because the associated codimension-$(p+1)$ symmetry defects can always be exchanged without crossing, leading to abelian fusion rules.} For instance, adding fields of electric charge~$N$ to Maxwell theory breaks~$U(1)_e^{(1)} \rightarrow {\mathbb Z}_N^{(1)}$. Another important example is the~${\mathbb Z}_N^{(1)}$ center symmetry of pure~$SU(N)$ Yang-Mills theory (and more generally, the center symmetry of any pure gauge theory).\footnote{~In these cases the reduction of the center symmetry to~${\mathbb Z}_N^{(1)}$ can be understood in terms of screening of the Wilson lines by electrically charged matter.} Due to their importance in gauge theory, we will mostly focus on 1-form symmetries, but the discussion holds for all~$p$-form symmetries with~$p \geq 1$. 

\item All 1-form symmetries~$G^{(1)}$ naturally act on line defects by linking them with the codimension-2 topological symmetry defects~$U^{(1)}(g, \Sigma_{d-2})$. For instance, the line operators charged under the~${\mathbb Z}_N^{(1)}$ center symmetry of~$SU(N)$ gauge theory are the Wilson lines. However, the 1-form symmetry defects cannot link local operators, and hence such  symmetries commute with local operators, as well as with the point particle states they create or with the~$S$-matrix of these particles. This is how~$(p \geq 1)$-form symmetries evade CM-like theorems. 

\item 1-form symmetries can spontaneously break. For instance, in dimension\footnote{~Similar, long familiar results apply in $d=3$, involving ordinary rather than higher-form symmetry: the 3d photon can be dualized to a compact scalar, which is the NG boson of spontaneously broken $U(1)_m^{(0)}$~\cite{Polyakov:1976fu}. In $d=3$, confinement arises either if $U(1)_m^{(0)}$ is restored by the vacuum, or if it is explicitly broken, e.g. by 3d instantons (4d monopoles). Explicit breaking gives a mass to the pseudo-NG boson scalar photon~\cite{Polyakov:1976fu,Affleck:1982as}.  For spontaneously broken $U(1)^{(p)}$ with $p>0$, the spin-$p$ NG boson does not get a mass from explicit breaking of the symmetry, see~\cite{McGreevy:2022oyu} and references cited therein for this {\it robustness} of higher-form symmetry. For $d=4$, the photon can be viewed as the NG boson of spontaneous breaking of either $U(1)_e^{(1)}$ or $U(1)_m^{(1)}$.}~$d \geq 4$ the photon of  theories with an abelian Coulomb phase,\footnote{~The low-energy theory has the IR-free photon of Maxwell theory.  The photon can have IR irrelevant, higher-derivative interactions if ~$U(1)_e^{(1)}$ is accidental; e.g. this happens for supersymmetric theories on their Coulomb branch.} is a NG boson for its spontaneously broken~$U(1)_e^{(1)}$ symmetry~\cite{Kovner:1992pu, Gaiotto:2014kfa, Hofman:2018lfz, Lake:2018dqm}. A spontaneously broken discrete 1-form symmetry leads to deconfined line defects (i.e.~defects obeying a perimeter law) at long distances. By contrast, unbroken 1-form symmetry implies confinement of the line operators charged under it.\footnote{~This extends Polyakov's description~\cite{Polyakov:1978vu}, making the thermal circle there inessential~\cite{Gaiotto:2014kfa}.} Thus, (un)broken 1-form symmetries characterize confined and deconfined phases in the spirit of the Landau paradigm.\footnote{~See~\cite{McGreevy:2022oyu} for a recent discussion of generalized symmetries through the lens of condensed matter physics.}  

\item A suitable background field for a~$G^{(1)}$ symmetry is a 2-form gauge field~$B^{(2)}$,\footnote{~If~$G^{(1)}$ is discrete, then~$B^{(2)}$ is valued in~$H^{2}(M_d, G^{(1)})$, where~$M_d$ is the spacetime manifold.} with its standard 1-form gauge transformations~$B^{(2)} \rightarrow B^{(2)} + d \Lambda^{(1)}$.\footnote{~Here~$\Lambda^{(1)}$ is the 1-form gauge parameter. Again a refinement is required if~$G^{(1)}$ is discrete.} As for ordinary symmetries, the theory is invariant under such transformations up to possible 't Hooft anomalies, which can be cancelled by a suitable invertible anomaly theory~${\cal A}_{d+1}[B^{(2)}]$. In the absence of 't Hooft anomalies, the~$G^{(1)}$ symmetry can be gauged by path-integrating over~$B^{(2)}$. 

\item Typically, gauging a~$p$-form symmetry~$G^{(p)}$ in~$d$ spacetime dimensions gives rise to a new theory with a~${\tilde G}^{(q)}$ global symmetry, whose degree~$q$ depends on whether~$G$ is discrete or continuous.\footnote{~More care is required if~$G^{(p)}$ has 't Hooft anomalies or if~$p =0$ and~$G$ is non-Abelian. See for instance~\cite{Bhardwaj:2017xup,Tachikawa:2017gyf}. } For instance, gauging a~${\mathbb Z}_n^{(p)}$ symmetry in~$d$ dimensions typically gives rise to a new theory with a~${\mathbb Z}_n^{(q = d - p - 2)}$ symmetry. This generalizes the familiar twisted sector (or quantum) symmetry of two-dimensional orbifolds, where~$d =2$ and~$p = 0$. 

\item More generally, one can gauge a discrete $p$-form symmetry only along a codimension-$q$ submanifold in spacetime. This is known as  higher gauging~\cite{Roumpedakis:2022aik}  of a $p$-form global symmetry. Higher gauging  does not change the bulk of the QFT, but generates a codimension-$q$ topological defect, known as a condensation defect \cite{Kong:2014qka,Else:2017yqj,Gaiotto:2019xmp,Kong:2020cie,Johnson-Freyd:2020twl,Roumpedakis:2022aik,Choi:2022zal}. 
Such condensation defects are generally not invertible, i.e.~they generate a non-invertible global symmetry (see section \ref{sec:noninv} below).

\end{itemize}

\subsection{Higher Group Symmetries}\label{sec:hg}

Higher group symmetries arise when~$p$-form symmetries~$G^{(p)}$ of different form degrees are present and can affect each other in various ways. Roughly speaking, if~$p < q$ then~$G^{(p)}$ can act on~$G^{(q)}$, while if~$p > q$ the background field~$B^{(p+1)}$ for~$G^{(p)}$ can shift under background gauge transformations of the background field~$B^{(q+1)}$ for~$G^{(q)}$. The precise form of the shift is determined by certain group cohomology (or Postnikov) classes.\footnote{~See for instance \cite{baez2004higher,Baez:2004in,Schreiber:2008} for a mathematical exposition of higher groups and related subjects.}

Higher-group gauge symmetry has a long history in string theory, most famously in the context of the Green-Schwarz (GS) mechanism and its generalizations. The fact that higher groups can also furnish global symmetries of QFTs is a more recent observation that is actively being explored and applied in many contexts, see~\cite{Kapustin:2013qsa,Kapustin:2013uxa,Gukov:2013zka,Thorngren:2015gtw,Sharpe:2015mja,Gaiotto:2017zba,Bhardwaj:2016clt,Tachikawa:2017gyf,Delcamp:2018wlb,Cordova:2018cvg,Benini:2018reh,Tanizaki:2019rbk,Bah:2020uev, Bhardwaj:2020phs,Hsin:2020nts,Cordova:2020tij,DelZotto:2020sop,Iqbal:2020lrt, Hidaka:2020izy, Brennan:2020ehu,Lee:2021crt, Apruzzi:2021vcu, Apruzzi:2021nmk,Bhardwaj:2021wif, Apruzzi:2021mlh,Hsin:2021qiy, DelZotto:2022joo,Cvetic:2022imb, Sharpe:2022ene, Pantev:2022kpl, DelZotto:2022fnw, DelZotto:2022ras} for an incomplete sampling of the recent literature.\footnote{~Note that higher-form global symmetries are a special case of higher-group global  symmetries.}  

Rather than discuss higher-group symmetries in general, we will focus on a concrete and elementary example: a 2-group symmetry involving~$G^{(0)} = U(1)^{(0)}_A$ and~$G^{(1)} = U(1)^{(1)}_B$.\footnote{~Here we assume that~$d > 2$. For comments on the~$d = 2$ case see for instance~\cite{Sharpe:2015mja}.} The relevant Postnikov class resides in~$H^3(G^{(0)}, G^{(1)})$, and we assume that it is non-vanishing.\footnote{~Such  examples can be engineered in 4d Abelian gauge theories with fermionic mater, where the Postnikov data is determined by certain mixed triangle anomaly coefficients involving two~$j_A^{(1)}$ currents and one abelian gauge field. In this case~$U(1)_B^{(1)}$ is the magnetic 1-form symmetry of that gauge field (assuming no magnetic matter).}
In this case the fusion of the~$U(1)_A^{(0)}$ current~$j_A^{(1)}$ with itself contains the~$U(1)_B^{(1)}$ current~$J_B^{(2)}$. Schematically,
\begin{equation}
j^{(1)}(x)_A j^{(1)}_A(0) \sim K(x) J_B^{(2)}(0) + \cdots~,
\end{equation}
where~$K(x)$ is a known, singular function depending on the Lorentz indices of the currents. Its overall normalization is determined by the Postnikov class. This fusion rule is reminiscent of the non-Abelian fusion rule for~$G^{(0)}$ currents above, except that it involves currents of different form degree.  The structure of this OPE forces us to deform the algebra of background gauge transformations. In particular, we find that the background field~$B^{(2)}$ that couples to~$J_B^{(2)}$ receives a background GS shift depending on the background field~$A^{(1)}$ that couples to~$j_A^{(1)}$, as well as on the Postnikov class. Additional details can be found in~\cite{Cordova:2018cvg,Cordova:2020tij}.

\subsection{Non-Invertible Symmetries}\label{sec:noninv}

As reviewed in Section \ref{sec:ord}, in relativistic systems, the modern characterization of an ordinary global symmetry is in terms of its topological  symmetry operators and defects of codimension one in spacetime. 
A natural question then arises: While every ordinary global symmetry element is associated with a topological defect, is the converse true?
This turns out \textit{not} to be the case: there are topological defects not associated with any ordinary global symmetry. 
These defects are \textit{non-invertible} because there is no inverse in the fusion algebra and their fusion can involve more than one operator. 

In 1+1d, non-invertible topological lines are ubiquitous, especially in rational CFTs and integrable field theories.\footnote{~The search for hidden symmetry has a long history in these theories. For instance,  in rational CFTs any primary operator $g$ with a fusion rule that only contains the identity operator on the right-hand side leads to an invertible abelian symmetry of the theory that can be used to extend the chiral algebra~\cite{Moore:1988qv,Intriligator:1989zw,Schellekens:1989am}.} In rational CFT \cite{Moore:1988qv,Moore:1989yh,Fuchs:2002cm}, there is  a general construction of topological lines that commute with the extended chiral algebra \cite{Verlinde:1988sn,Petkova:2000ip}.  
The simplest example is the Kramers-Wannier duality line in the Ising CFT \cite{Frohlich:2004ef,Frohlich:2006ch,Frohlich:2009gb}. 
The fusion of two non-invertible duality lines $\cal N$ gives the sum of the identity line and the $\mathbb{Z}_2$ symmetry line $\eta$:
\begin{align}
{\cal N}\times {\cal N}=1+\eta\,.
\end{align}
As we bring the (local) order operator past this line, it  becomes a (non-local) disorder operator on the other side.

In recent years, it has been advocated that these non-invertible topological defects should be viewed as  a generalization of  ordinary global symmetries \cite{Bhardwaj:2017xup,Chang:2018iay}. 
The fusion and crossing properties of these topological lines are mathematically captured by a fusion category.
The generalized gauging  of these lines in 1+1d has been developed in \cite{Frohlich:2009gb,Carqueville:2012dk,Brunner:2013xna,Bhardwaj:2017xup,Komargodski:2020mxz,Gaiotto:2020iye,Huang:2021zvu}. 
When there is an obstruction to gauging, the corresponding anomalies lead to non-trivial dynamical constraints on the renormalization group flows \cite{Chang:2018iay,Thorngren:2019iar,Komargodski:2020mxz,Thorngren:2021yso}. 
The relation to fermionic anomalies has been discussed in \cite{Thorngren:2018bhj,Ji:2019ugf,Lin:2019hks,Kaidi:2021xfk,Burbano:2021loy}.

In higher spacetime dimensions, there are also  examples of non-invertible defects of various codimensions. 
In 2+1d, the non-abelian anyons in a low-energy TQFT, described by the theory of modular tensor categories,  can be viewed as the non-invertible generalization of one-form global symmetries (see \cite{Kaidi:2021gbs,Buican:2021axn,Yu:2021zmu,Benini:2022hzx,Roumpedakis:2022aik} for recent discussions). 
Recently, the non-invertible Kramer-Wannier duality defects have been generalized to 3+1d gauge theories \cite{Koide:2021zxj,Choi:2021kmx,Kaidi:2021xfk,Hayashi:2022fkw,Choi:2022zal,Bhardwaj:2022yxj,Kaidi:2022uux} via the gauging of  higher-form global symmetries. 
More generally, in any theory with a higher-form symmetry,  the most fundamental non-invertible symmetries are condensation defects \cite{Kong:2014qka,Else:2017yqj,Gaiotto:2019xmp,Kong:2020cie,Johnson-Freyd:2020twl} arising from \textit{higher gauging}   \cite{Roumpedakis:2022aik}. 
Condensation defects in 2+1d were systematically analyzed in\cite{Roumpedakis:2022aik}, generalizing the results of \cite{Fuchs:2002cm,Kapustin:2010if}.    
The non-invertible fusion ``coefficients" of these $n$-dimensional topological defects are generally not numbers, but $n$-dimensional TQFTs \cite{Roumpedakis:2022aik,Choi:2022zal}.  
More recently, infinitely many non-invertible global symmetries were identified in the Standard Model and axion models from  ABJ anomalies \cite{Choi:2022jqy,Cordova:2022ieu}.

Non-invertible defects have also found applications in quantum gravity in the context of the completeness hypothesis  \cite{Rudelius:2020orz,Heidenreich:2021xpr,McNamara:2021cuo, Cordova:2022rer, Arias-Tamargo:2022nlf}. They have also been realized on the lattice \cite{Grimm:1992ni,Feiguin:2006ydp,Hauru:2015abi,Aasen:2016dop,Buican:2017rxc,Aasen:2020jwb,Inamura:2021szw,Koide:2021zxj,Huang:2021nvb,Vanhove:2021zop,Liu:2022qwn}, sometimes under the name of ``algebraic higher symmetry'' \cite{Ji:2019jhk,Kong:2020cie}.

\subsection{Subsystem Symmetries and Fractons}\label{sec:fractons}

A novel kind of global symmetry, known as \textit{subsystem global symmetry}, has featured prominently in many condensed matter systems, including the gapless model of \cite{PhysRevB.66.054526} and many gapped fracton models \cite{PhysRevLett.94.040402,PhysRevA.83.042330,Vijay:2016phm}.
(See \cite{Nandkishore:2018sel,Pretko:2020cko,Grosvenor:2021hkn} for reviews on fractons, and \cite{Brauner:2022rvf} for a Snowmass white paper on field theories for fractons.)

Subsystem  symmetry is similar to the higher-form  symmetry, but they differ in important ways (see  \cite{Seiberg:2019vrp,Qi:2020jrf} for related discussions). 
For both kinds of global symmetries, the conserved charges are supported on closed,  higher-codimensional manifolds $\mathcal{S}$ in space.  
Unlike the higher-form symmetry charge, the charge of a subsystem  symmetry depends not only on the topology of $\mathcal{S}$, but possibly also on the shape and the location of $\mathcal{S}$.  
Furthermore, the charge might only  be allowed to be on certain $\mathcal{S}$ but not all  manifolds of a given codimension. 
On a lattice, the number of independent conserved charges grows  with the number of sites, leading to a huge global symmetry in the continuum limit.


Just like ordinary global symmetries, subsystem global symmetries can be spontaneously broken \cite{Batista:2004sc,2017EPJST.226..749J,paper1,paper2,paper3,Gorantla:2021bda,Distler:2021qzc}.  
They can also have 't Hooft anomalies \cite{paper3,Gorantla:2021svj,Burnell:2021reh}, which are obstructions to gauging them.  These anomalies are captured by Subsystem Symmetry Protected Topological (SSPT) phases in one higher dimension \cite{Raussendorf2019,You2018,You:2018zhj,Devakul:2018fhz,Stephen2019subsystem,Devakul:2019duj,Burnell:2021reh,Yamaguchi:2021xeq}. 
The symmetry fractionalization of subsystem symmetries has been recently discussed in \cite{Stephen:2022tjg}.

Fractons are a large class of lattice models with  subsystem global and gauge symmetries.
In 3+1 dimensions, fracton models are robust topological phases  with many peculiar properties, including infinite ground state degeneracy that is sub-extensive in the system size,  and excitations with restricted mobility.  
Because of these unusual  characteristics, gapped fracton models cannot be described by conventional TQFT. 
Some fracton models can be realized by  gauging a subsystem symmetry\cite{Vijay:2016phm,Williamson:2016jiq,Slagle:2017wrc,Shirley:2018vtc,Gromov:2018nbv,paper3,Gorantla:2020jpy}.  
In the low-energy limit, they are typically described  by certain exotic higher-rank tensor gauge theories \cite{Gu:2006vw,Xu:2006,Pankov:2007,Xu2008,Gu:2009jh,rasmussen,Pretko:2016kxt,Pretko:2016lgv,Prem:2017kxc,Slagle:2017wrc,2018PhRvL.121w5301P,2018PhRvB..98l5105M,Bulmash:2018lid,2018PhRvB..98c5111M,Slagle:2018kqf,Pretko:2018jbi,Williamson:2018ofu,2020PhRvR...2b3249Y,Seiberg:2019vrp,You:2019bvu,Gromov:2020rtl,Gromov:2020yoc,paper1,paper2,paper3,Slagle:2020ugk,Karch:2020yuy,Yamaguchi:2021qrx,Geng:2021cmq,Hsin:2021mjn,Du:2021pbc,Bidussi:2021nmp,Jain:2021ibh,Gorantla:2022eem,Perez:2022kax}.


The most significant consequence of subsystem  symmetry   is that  the low-energy observables  are  sensitive to details of the short-distance physics. 
For instance, in fracton models the large ground state degeneracy is enforced by the subsystem global symmetries and their 't Hooft anomalies~\cite{paper3,Burnell:2021reh}. 
Similarly, the restricted mobility of the fracton excitations is traced to the selection rules arising from a subsystem global symmetry, the \textit{time-like global symmetry}, that acts on defects, rather than operators \cite{Gorantla:2022eem}. 
This surprising UV/IR mixing \cite{paper1,Gorantla:2021bda,You:2021tmm,Zhou:2021wsv,You:2021sou}, which is reminiscent of that in quantum field theory on a non-commutative space \cite{Minwalla:1999px},  is the main reason why these exotic systems  defy a conventional continuum field theory description \cite{Slagle:2017wrc,Bulmash:2018knk,Slagle:2018swq,paper1,paper2,paper3,Gorantla:2020xap,Aasen:2020zru,Slagle:2020ugk,Gorantla:2020jpy,Rudelius:2020kta,Gorantla:2021svj,Hsin:2021mjn,Gorantla:2021bda,Distler:2021bop,Lake:2021pdn}.

A summary of the various kinds of symmetry described above is given in Table \ref{table:sym}.
 
 \begin{table}[h]
\begin{align*}
\left.\begin{array}{|c|c|c|c|c|}
\hline \text{Properties of} &~~ \text{Ordinary} ~~&~~ \text{Higher-form}~~&~~ \text{Non-invertible}~~  & \text{Subsystem} \\
~~\text{symmetry operator}~~&\text{symmetry}&\text{symmetry}&\text{symmetry}&\text{symmetry}\\
\hline \text{Codimension} & 1 & >1 & \ge 1 & >1 \\
\text{in spacetime}&&&&\\
\hline \text{Topological} & \text{yes} & \text{yes} &\text{yes}&  \text{not completely} \\
&&&&\text{\small but conserved in time}\\
\hline \text{Fusion rule} & \text{group} & \text{group} & \text{category}  & \text{group}\\
&&&&\\
\hline \end{array}\right.
\end{align*}
\caption{Properties of symmetry defects for various kinds of global symmetry~\cite{review}. }\label{table:sym}
\end{table}

 \section{Anomalies and Dynamical Consequences of Symmetry}\label{sec:anomdyn}
 
 \subsection{Anomalies and Anomaly Matching} \label{sec:anomalies}

 The strong force, confinement,  dualities, and emergent degrees of freedom illustrate how renormalization group (RG) flows can lead to surprising interconnections between classically very different UV descriptions of IR physics, or between the UV and IR limits of RG flows.    A key feature 
 of any concept of symmetry is how it survives continuous, symmetry preserving deformations of the theory.  This includes in particular renormalization group flows: if at short distances a model supports a given symmetry, then the same symmetry must be present in any long distance effective field theory.\footnote{~Of course symmetries can be enhanced in the IR, by additional, accidental components. Moreover, symmetries can decouple in the IR if they only act non-trivially on degrees of freedom that have been decoupled from the low-energy theory. Most precisely, there is a homomorphism from the symmetry group of the UV to that of the IR.}
 
 This idea of global symmetry matching between the ultraviolet and the infrared becomes remarkably powerful when extended to include 't Hooft anomalies of global symmetries.  
 't Hooft's pioneering idea~\cite{tHooft:1979rat} was to view such anomalies as an obstruction to gauging a global symmetry, and to argue that  they must be invariant under the renormalization group by coupling to a suitable set of spectator (i.e.\ weakly interacting) fermions.  Subsequently, this idea of anomaly matching was understood as constraints on various structure functions of conserved current correlation functions \cite{Frishman:1980dq,Coleman:1982yg}.  
 
 The modern point of view of anomalies is that of anomaly inflow \cite{Callan:1984sa} (see also~\cite{Faddeev:1984ung}) where an anomaly of a $d$-dimensional quantum field theory with spacetime $X_d$ is presented via a theory  ${\cal A}_{d+1}$ in $d+1$ spacetime dimensions, with bulk spacetime $Y_{d+1}$ such that the original spacetime is the boundary of the bulk spacetime, $X_d=\partial Y_{d+1}$.    In some contexts, e.g. \cite{Callan:1984sa, Freed:1998tg, Harvey:2005it}, the $d+1$-dimensional anomaly theory ${\cal A}_{d+1}$ has a clear physical interpretation, with the  lower dimensional anomalous theory arising as a defect in the ambient physical system.\footnote{~If the anomalous symmetries are gauged on the defect, the bulk theory is physical and cannot be decoupled;  there is no ``genuine" $d$ dimensional theory in this case.   For 't Hooft anomalies, the dynamics of the bulk $d+1$ dimensional theory can decoupled, leaving only the anomaly. A classic example are  the $d+1$ dimensional Wess-Zumino-Witten terms for the pions of 4d QCD in the chiral symmetry broken phase \cite{Witten:1983tw,Witten:1983tx} and 2d non-linear sigma models \cite{Witten:1983ar}.  Another example are the chiral, domain wall fermions of~\cite{Kaplan:1992bt}, where the defect arises via spatially dependent backgrounds (the Dirac mass) in the bulk.}
 This point of view is conceptually similar to that of symmetry protected topological phases in condensed matter physics, where a trivially gapped system in bulk gives rise to anomalous edge modes on its boundary.  From a more abstract perspective, the anomaly theory generalizes the concept of projective representations, which describe simple anomalies in quantum mechanics (see e.g.\ \cite{Gaiotto:2017yup}), to the quantum field theory setting. The anomaly theory can be taken as the definition of the anomaly, and this notation can be applied to generalized symmetries. The anomaly theory ${\cal A}_{d+1}$ has been significantly developed and applied in recent years in the high energy, condensed matter, and mathematics communities. This includes results for familiar continuous global symmetries \cite{Thorngren:2014pza, Ohmori:2014kda, Wang:2018qoy, Lee:2020ojw, Lee:2020ewl,Anber:2020gig, Lee:2022spd}, discrete global symmetries \cite{Kapustin:2014lwa, Kapustin:2014zva, Kapustin:2014dxa, Wang:2014pma, Gaiotto:2015zta, Bhardwaj:2016clt, Guo:2017xex, Guo:2018vij,Tong:2019bbk, Grigoletto:2021zyv, Smith:2021vbf}, higher-form and higher-group symmetries~\cite{Kapustin:2013uxa, Thorngren:2015gtw, Cordova:2018cvg, Benini:2018reh, Hsieh:2020jpj, Hsin:2021qiy}, and subsystem symmetries \cite{paper3,Gorantla:2021svj,Burnell:2021reh}, families of quantum field theories \cite{Kikuchi:2017pcp, Thorngren:2017vzn, Tachikawa:2017aux, Tanizaki:2018xto, Seiberg:2018ntt, Hsieh:2019iba, Cordova:2019jnf, Cordova:2019uob, Hsin:2020cgg, Choi:2022odr}, as well as structural developments \cite{Kapustin:2014tfa, Kapustin:2014gma, Seiberg:2016rsg, Tachikawa:2017gyf, Thorngren:2018bhj,  Wan:2018bns, Cordova:2019wpi, Yonekura:2020upo, Hason:2020yqf, Thorngren:2020yht, Gaiotto:2017zba, Delmastro:2021xox, Cherman:2021nox}.

 For continuous global symmetries in even $d$, the anomaly theory ${\cal A}_{d+1}$ is often a Chern-Simons-type  TQFT, as in the descent formalism of \cite{Alvarez-Gaume:1984zlq} and references therein.  More generally, global or non-perturbative anomalies can be described via the $\eta$-invariant~\cite{Dai:1994kq} which can also be described via inflow~\cite{Yonekura:2016wuc, Witten:2019bou}.  In the context of $AdS_{d+1} \leftrightarrow CFT_d$ duality, the anomaly theory ${\cal A}_{d+1}$ is a topological subsector of the $AdS_{d+1}$ bulk gravity theory \cite{Witten:1998qj}.      The anomaly theory ${\cal A}_{d+1}$ for a traditional QFT merely encodes an overall phase, so it is associated with a one-dimensional Hilbert space. The anomaly theory in such cases is called {\it invertible}~\cite{Freed:2014iua, Freed:2014eja} (see also \cite{Gaiotto:2017zba, Yonekura:2018ufj,Yamashita:2021cao}).\footnote{~There are also more exotic $d$-dimensional QFTs, sometimes called relative~\cite{Freed:2012bs} or meta or non-genuine theories, where the ${\cal A}_{d+1}$ theory plays a more essential role than just a projective phase and is instead associated with a multi-dimensional Hilbert space.  This was seen in the context of 2d chiral RCFTs, where the bulk is a 3d Chern-Simons theory~\cite{Moore:1988qv, Elitzur:1989nr}. Other examples include certain 6d ${\cal N}=(2,0)$ theories,  which arise at the boundary of 7d Chern-Simons type theories, see e.g.~\cite{Witten:1998wy,Gaiotto:2014kfa,Monnier:2017klz,Gukov:2020btk}.}
  
 \subsection{Applications to QFT Dynamics} \label{sec:dynamics}

There are a wide variety of applications of anomalies and symmetry to the dynamics of quantum field theory.  Due to their robust nature, symmetries and their associated anomalies can often be reliably computed in weakly coupled limits of a theory, and then applied to constrain the dynamics of strongly coupled phases.  For ordinary continuous global symmetries,   non-zero 't Hooft anomalies preclude a mass gap: the IR theory must have massless chiral fields that saturate the anomalies in the unbroken case, or massless NGBs (with a WZW-type term) in the spontaneously broken case.  For discrete global symmetries, non-zero 't Hooft anomalies can also in some cases be matched by a gapped TQFT (if an obstruction~\cite{Cordova:2019bsd, Cordova:2019jqi} vanishes).  't Hooft anomaly matching 
played a key role in the discovery of electromagnetic dualities in four-dimensional gauge theories  \cite{Seiberg:1994pq, Intriligator:1995id, Intriligator:1995au} and has since become an essential field theory technique.  Modern applications are inspired by this example but expand the logic to the more general concepts of symmetry described above, and hence to the more general class of invertible field theories encoding their anomalies.  Below we highlight several classes of examples.

Higher form symmetry has been crucial in understanding the infrared behavior of Chern-Simons matter theories in three dimensions.  In this context one also finds a close link to fundamental concepts in three-dimensional topological field theory where one-form symmetry generators can be interpreted as abelian anyons \cite{Kapustin:2014gua, Gaiotto:2014kfa, Tachikawa:2016cha, Tachikawa:2016nmo, Burnell:2017otf, Hsin:2016blu, Hsin:2018vcg, Lee:2018eqa, Delmastro:2019vnj}.  The language of higher form symmetry then gives a way to understand the properties of these line defects in a broader context of non-topological field theories.  In particular this has given rise to an explosion of new level-rank and boson-fermion dualities for such interacting phases \cite{Aharony:2015mjs, Hsin:2016blu, Aharony:2016jvv, Benini:2017dus, Cordova:2017vab, Benini:2017aed, Cordova:2017kue, Benini:2018bhk, Cordova:2018qvg, Bashmakov:2018wts}, as well as new results for the phase diagrams of strongly interacting gauge theories \cite{Freed:2017rlk, Komargodski:2017keh, Komargodski:2017dmc, Gomis:2017ixy, Cordova:2017vab, Cordova:2017kue, Komargodski:2017smk, Choi:2018tuh, Cordova:2018qvg,Armoni:2019lgb}.

In four spacetime dimensions there have also been extensive recent applications of symmetry to constrain the dynamics of gauge theories \cite{Gukov:2013zka, Tachikawa:2016nmo, Putrov:2016qdo, Shimizu:2017asf, Cherman:2017tey, Garcia-Etxebarria:2018ajm, Aitken:2018kky, Cordova:2018acb,Wan:2018zql,  Tanizaki:2019rbk, Yonekura:2019vyz, Wang:2019obe, Wan:2019soo, Wan:2019oax,  Wan:2020ynf, Cherman:2020hbe, DelZotto:2020esg,Hsin:2020nts, Poppitz:2020tto, Anber:2020qzb, Razamat:2020kyf, Anber:2021lzb, Anber:2021iip, Anber:2021upc, Lee:2021crt, Unsal:2021cch,Bhardwaj:2021wif, Cox:2021vsa,  Anber:2021upc, Bhardwaj:2021mzl, Wang:2021ayd, Tanizaki:2022ngt, Wang:2022ucy}.  In these gauge theory contexts, one-form symmetry finds a natural utilization as the symmetry governing the transition between confinement and deconfinement \cite{Gaiotto:2014kfa}.  A highlight is the identification of an anomaly at $\theta=\pi$ in certain pure Yang-Mills theories implying non-trivial long-distance dynamics of these models \cite{Gaiotto:2017yup}.  There is also a natural link to the three-dimensional Chern-Simons matter results which often manifest on domain walls in four dimensions \cite{Gaiotto:2017tne, Bashmakov:2018ghn, Anber:2018xek,Cox:2019aji, Anber:2020xfk, Delmastro:2020dkz}.  Further there is a sharp interplay between anomalies, spontaneous symmetry breaking, and the mass gap in such theories \cite{Garcia-Etxebarria:2017crf,Wang:2017loc, Wan:2018djl, Wang:2018edf,Kobayashi:2019lep, Cordova:2019bsd, Cordova:2019jqi, Thorngren:2020aph}.

The extensive appearance of higher-form symmetry in quantum field theory means that frequently one finds higher group global symetries, which have their own anomalies and dynamical implications.  Higher-group global symmetry has also been extensively explored in five and six-dimensional quantum field theory and little string theories \cite{Albertini:2020mdx, Gukov:2020btk, Bah:2020uev, Morrison:2020ool, Apruzzi:2021phx, Apruzzi:2021nmk, Hubner:2022kxr, DelZotto:2022joo, Cvetic:2022imb, DelZotto:2022fnw,  Heckman:2022suy}, where in the latter continuous higher symmetry is particularly natural due to the presence of a conserved string number \cite{Cordova:2016emh, Cordova:2020tij}.  In particular, these ideas have been applied to better understand certain supersymmetric dualities \cite{Bah:2018gwc,Bah:2018jrv, Bah:2020jas, DelZotto:2020sop, Bhardwaj:2021pfz, Bhardwaj:2021zrt, Bhardwaj:2021ojs}, and to derive universal constraints on renormalization group flows \cite{Cordova:2015fha,Heckman:2015axa,Cordova:2019wns}.

Similar to the anomaly constraints from an ordinary symmetry, the non-invertible  symmetries also have dramatic consequences on renormalization group flows. 
In two-dimensional QFT,  invertible and non-invertible topological lines have been explored extensively in \cite{Bhardwaj:2017xup,Chang:2018iay,Lin:2019kpn,Thorngren:2019iar,Cordova:2019wpi,Lin:2019hks,Komargodski:2020mxz,Yu:2020twi,Chang:2020imq,Pal:2020wwd,Nguyen:2021naa, Sharpe:2021srf, Hegde:2021sdm,Lin:2021udi,Inamura:2021wuo,Grigoletto:2021zyv,Thorngren:2021yso,Huang:2021ytb,Huang:2021zvu, Gukov:2021swm} in recent years with various dynamical applications.  
In particular, using a modular invariance argument, it was shown in \cite{Chang:2018iay} (see also \cite{Thorngren:2019iar} for generalizations) that the existence of certain non-invertible topological lines is incompatible with a trivially gapped phase. There are also recent applications to  2d adjoint QCD, clarifying the infrared behavior \cite{Komargodski:2020mxz}.  In four-dimensional theories the dynamical implications of non-invertible symmetry are just beginning to emerge with recent results on anomalies of such symmetries \cite{Choi:2021kmx, Choi:2022zal} and applications to Yang-Mills theory \cite{Kaidi:2021xfk}.


\section*{Acknowledgements}

We are grateful to our collaborators and colleagues who have helped shape this exciting subject, and to the organizers of the formal theory Snowmass process, L. Rastelli, and D. Poland, for inviting us to contribute this white paper. CC, TD, and KI acknowledge support from the Simons Collaboration on Global Categorical Symmetries. In addition CC is supported by the US Department of Energy DE-SC0021432; TD is supported by a DOE Early Career Award under DE-SC0020421 and the Mani L. Bhaumik Presidential Chair in Theoretical Physics at UCLA; KI is supported by DOE award DE-SC0009919 and Simons Foundation award 568420. 

\bibliography{ref}
 
\bibliographystyle{JHEP}

\end{document}